\documentstyle[12pt]{article}

\begin{document}
\title{Fields with continuously distributed mass}
\author{N.V.Krasnikov  
  \\
INR RAN, Moscow 117312}

\maketitle
\begin{abstract}

We describe
local field theories with continuously 
distributed mass. Such models 
can be realized as models in $d  > 4 $ space-time with Poincare invariance only in 
four-dimensional space-time. We also discuss some possible phenomenological consequences.
Namely, we show that the  Higgs boson phenomenology in the SM extension  with continuously 
distributed Higgs boson mass can differ  
in a drastic way from  the standard Higgs boson phenomenology,

\end{abstract}

\newpage
\section{Introduction}

In this report based on refs.\cite{1}-\cite{3}\footnote{See also \cite{4}-\cite{6}.} we discuss 
local field theories with continuously 
distributed mass. We show that such models could be renormalizable. Moreover such models 
can be realized as models in $d  > 4 $ space-time with Poincare invariance  only in 
four-dimensional space-time. We also discuss some possible phenomenological consequences.
Namely, we show that the  Higgs boson phenomenology in the SM extension  with continuously 
distributed Higgs boson mass can differ  in a drastic way from  the standard Higgs boson phenomenology, 
We also point out  that the notion of an unparticle, 
 introduced by Georgi \cite{7}  can be interpreted as a 
particular case of a field  with continuously distributed mass.

\section{Some examples}

Let us start with N  scalar fields 
$\phi_k(x)$
with masses $m_k ~(k = 1,2,...N)$.
For the field 
$\phi(x, m_k,c_k, N) = \sum_{k = 1}^{N}c_k\phi_k(x)$  free propagator
 has the form 
\begin{equation}
D(p^2, m_k, c_k,N) = \sum_{k = 1}^{N} \frac{|c_k|^2}
{(p^2 - m^2 +i\epsilon)} = \int _0^{\infty}\frac{\rho(t, c_k, m_k, N)}
{p^2 -t+i\epsilon}dt   \,,
\end{equation}
where the spectral density is $\rho(t, c_k, m_k, N) = 
 \sum_{k =1}^{N} |c_k|^2 \delta(t - m^2_k)$.
In the limit $ N \rightarrow \infty$ 
$\rho(t, c_k, m_k, N) \rightarrow \rho(t)$ and the propagator 
 $ D(p^2, m_k, c_k,N) \rightarrow D(p^2) =    \int _0^{\infty}
\frac{\rho(t)}{p^2 -t+i\epsilon}dt$.
For instance, for $m^2_k = m^2_0 +  \frac{k}{N} \Delta^2  $ and 
$|c_k|^2 = \frac{1}{N}$ we find that the limiting spectral 
density is $\rho(t) = \frac{1}{\Delta^2}
\theta (t - m^2) \theta(m^2 + \Delta^2 -t)$. 
For the limiting spectral density $ \rho(t) \sim t^{\delta-1}$ we find that 
    $D(p^2) \sim (p^2)^{\delta - 1}$ that corresponds to the 
case of unparticle propagator. In other words, for the limiting 
spectral density  $ \rho(t) \sim t^{{\delta}-1}$ the field 
$\phi(x, \rho(t))  $  can be interpreted as unparticle.
\footnote{The interpretation of the unparticle as a tower of massive 
particles 
was also proposed in ref.\cite{8}}   
One can introduce 
the selfinteraction Lagrangian in 
standard way as 
\begin{equation}
L_{int}(\phi(x,\rho(t)) = -\lambda (\phi(x,\rho(t)))^4 \,.
\end{equation}
For finite  $\int _0^{\infty}\rho(t)dt $
the asymptotics of  propagator  $D(p^2) \sim \frac{1}{p^2}$ 
and the model (2)  is renormalizable. It should 
be noted that for Georgi noninteracting scalar unparticle 
the effective Lagrangian has the form 
\begin{equation}
L_{unp} = \frac{1}{2}\partial_{\mu} \phi (-\frac{\partial^{\mu}
\partial_{\mu}}{M^2})^{-\delta}  \partial^{\mu}\phi \,.
\end{equation}

The fields with continuously distributed mass arise 
naturally in d-dimensional field theories.
Consider  five-dimensional scalar field  with the Lagrangian
\begin{equation}
L_5 = \frac{1}{2}(\partial_{\mu} \phi \partial^{\mu}\phi - 
\phi f(-\partial_4^2)\phi) \,,
\end{equation}
where $\mu = 0,1,2,3$. 
The Lagrangian (4) is invariant only under the four-dimensional Poincare group and 
for arbitrary $f(-\partial^2_4)$
it is not invariant under five-dimensional Poincare group.\footnote{The Lagrangian (4) is 
invariant under 5-dimensional Poincare-group for $ f(-\partial_4^2) =  -\partial_4^2 $.}
For the Lagrangian $L_5$ free propagator has the form 
\begin{equation}
D_0 = \frac{1}{p_{\mu}p^{\mu} - f(p^2_4)} \,.
\end{equation}
For the field $\phi(x, x_4 = 0)$ 
propagator is proportional to $\frac{1}{2\pi}\int^{\infty}_{-\infty}\frac{dp_4}{
p_{\mu}p^{\mu} - f(p^2_4) + i\epsilon}$  that corresponds to the case of the 
field with continuously distributed mass. 
Usually in the literature only models in $d>4$ space-time with d-dimensional Poincare group invariant 
Lagrangians are considered. 
However it should be stressed that from the experimental point of view 
we have to postulate the invariance of the Lagrangian only  under the four-dimensional Poincare group. 
For the interaction 
\begin{equation}
L_I = -\lambda\phi^4
\end{equation} 
and for some functions $f(p^2_4)$ the model is 
renormalizable\footnote{For instance, for $ f(p^2_4) = m^2$ for $|p_4| \leq p_0$ and 
$f(p^2_4) =\infty$ for $|p_4| > p_0$.}. Remember that standard  $\phi^4$-model 
is nonrenormalizable in $d>4$ space-time.  

It is possible to construct models where some fields exist in four-dimensional space-time (four-dimensional brane) 
and other fields live in  $(d>4)$ space-time. One of the simplest examples is  the model \cite{1} where scalar field 
$\phi(x, x_4)$ propagates in five-dimensional space-time and interacts with 
the four-dimensional fermion  field $\psi(x)$. 
The action of the model has the form
\begin{equation}
S_t = S_1 + S_2 + S_i \,,
\end{equation}
where 
\begin{equation}
S_1 = \int \frac{1}{2}[\partial_{\mu}\phi(x,x_4)\partial^{\mu}\phi(x,x_4) -\phi(x,x_4)f(-\partial^2_4)
\phi(x,x_4)]d^5x \,,
\end{equation} 
\begin{equation}
S_2 = \int \bar{\psi}(x)[i\gamma^{\mu}\partial_{\mu}]\psi(x)d^4x\,,
\end{equation}
\begin{equation}
S_i = \int [h\phi(x, x_4 =0)\bar{\psi(x)}\psi(x) -\lambda\phi^4(x,x_4=0)]d^4x \,,
\end{equation}
where $ x = (x_0,x_1,x_2,x_3) $, $ \partial_4 = \partial / \partial x_4 $, 
$d^5x = d^4xdx_4$, $ d^4x = dx_0 dx_1 dx_2 dx_3 $. 
It should be stressed 
that the model (7-10) is a local one in four-dimensional 
space-time. The Feynman rules for the model (7-10) coincide with the Feynman rules for the four-dimensional Yukawa model,
 the single difference is that instead of the free propagator $(p^2 - m^2 + i\epsilon)^{-1} $
  for the standard four-dimensional  scalar field we have to use the effective propagator 
\begin{equation}
D^{eff}(p^2) = (2\pi)^{-1} \int [p^2 - f(p^2_4) + i\epsilon]^{-1}dp_4
\end{equation}
for the four-dimensional field $\phi(x,x_4 = 0)$.

There are many generalizations  to the case of vector fields. For instance, consider 
the Stueckelberg Lagrangian
\begin{equation}
L_0 = \sum_{k=1}^{N}[-\frac{1}{4e^2_k}F^{\mu\nu,k}F_{\mu\nu,k} 
+\frac{m^2_k}{2e^2_k}(A_{\mu,k} - \partial_{\nu}\phi_k)^2] \,,
\end{equation}
where $F_{\mu\nu,k} = \partial_{\mu}A_{\nu,k} - \partial_{\nu}A_{\mu,k}$. 
   The Lagrangian (12) is invariant under gauge transformations 
\begin{equation}
A_{\mu,k} \rightarrow A_{\mu,k} +\partial_{\mu}\alpha_k \,,
\end{equation}
\begin{equation}
\phi_k \rightarrow \phi_k + \alpha_k \,.
\end{equation}
For the field $B_{\mu} = \sum_{k=1}^N A_{\mu,k}$ free propagator in 
transverse gauge is 
\begin{equation}
D_{\mu\nu}(p) = (g_{\mu\nu} - \frac{p_{\mu}p_{\nu}}{p^2})
(\sum^N_{k=1}( \frac{e^2_k}{p^2-m^2_k})) \,. 
\end{equation}
In the limit $N \rightarrow \infty $
\begin{equation}
D_{\mu\nu}(p)  
\rightarrow (g_{\mu\nu} - \frac{p_{\mu}p_{\nu}}{p^2}) D(p^2)\,, 
\end{equation}
where 
\begin{equation}
D(p^2)  = \int _0^{\infty}\frac{\rho(t)}{p^2 - t +i\epsilon}dt
\end{equation}
and $\rho(t) \geq 0$. 
One can introduce the interaction of the field $B_{\mu}$ with  
fermion field $\psi$ in standard way, namely 
\begin{equation}
L_{int} = \bar{\psi}\gamma_{\mu}\psi B^{\mu} \,.
\end{equation}
The simplest generalization of the SM  model  
consists in the the replacement of the $U(1)$ gauge 
field propagator 
\begin{equation}
  (g_{\mu\nu} - \frac{p_{\mu}p_{\nu}}{p^2})\frac{1}{p^2} \rightarrow 
(g_{\mu\nu} - \frac{p_{\mu}p_{\nu}}{p^2})D(p^2) \,.
\end{equation}
This generalization  preserves the renormalizability for finite  
$ \int _0^{\infty}\rho(t)dt$ because the ultraviolet asymptotics of 
$ D(p^2)$ coincides with  free propagator. 
For $\rho(t) \sim t^{\delta -1}$ we reproduce the case of vector unparticle.

\section{Possible phenomenological consequences}

Consider the SM in unitary gauge and instead of free propagator 
$(p^2 -m^2_H)^{-1} $ let us use the  propagator $D(p^2) = \int_{0}^{\infty}\rho(t)(p^2 -t)^{-1}dt$.
For finite $\int_0^{\infty}\rho(t)dt$ the model is renormalizable. For Breit-Wigner spectral 
density\footnote{For the spectral density (20) $\int \rho_{BW}(t)dt = 1$, 
$lim_{\Gamma \rightarrow 0} \rho_{BW}(t) = \delta(t - m^2_H)$.} 
\begin{equation}
\rho_{BW}(t) = (\frac{1}{\pi})\Gamma m_H[(t - m^2_H)^2 + \Gamma^2m^2_H]^{-1}
\end{equation}
one can interpret $\Gamma$ as an internal decay width of the Higgs boson \cite{1}.
For $\Gamma \gg \Gamma_{t}$, where $\Gamma_{t}$ is the standard Higgs boson decay width, 
the  Higgs boson will decay mainly into invisible modes that 
makes the Higgs boson discovery at the LHC extremely difficult.

Similar gauge invariant generalization of the SM is the following. Let us add to the SM fields  
 $SU_c(3)\otimes SU_L(2)\otimes U(1)$  
singlet scalar field $\phi(x, \rho(t))$ with continuously distributed mass. The interaction of 
the field   $\phi(x, \rho(t))$ with the Higgs doublet field $H(x)$ has the form  
\begin{equation}
L_{int}(\phi_{int}(x,\rho(t)), H(x)) = -\lambda_2 
(\phi_{int}(x,\rho(t))H^{+}(x)H(x) \,.
\end{equation}
After electroweak symmetry breaking the singlet field  
$\phi_{int}(x, \rho(t))$ will mix with the standard Higgs boson.
As a result of the mixing the Higgs boson  will have invisible decay 
modes as in previous example.

Another example  is  the $Z^{'}$ vector boson model  with 
continuously distributed mass. One of the possible effects due to nonzero internal decay width of the $Z^{'}$ 
boson is 
the existence of rather broad resonance structure in Drell-Yan reaction 
$pp \rightarrow Z^{'}+... \rightarrow l^{+}l^{-}+...$ . 

\section{Conclusion}
In this report we described quantum field theories with continuously distributed masses. It is 
possible to interpret  such models as quantum field theory models in $d >4$ space-time. 
The most interesting example is the Higgs boson with continuously distributed mass. The Higgs boson 
phenomenology for such model for $\Gamma \gg \Gamma_t$ is different from the standard Higgs boson 
phenomenology, namely, Higgs boson decays mainly into invisible modes that makes 
the LHC Higgs boson discovery  very untrivial.

This work was supported by the Grants   RFBR N07-02-00256, RFBR 08-02-91007-CERN.


\newpage
\begin{thebibliography}{99}

\bibitem{1} N.V.Krasnikov, Phys.Lett. {\bf B325} 430 (1994).
\bibitem{2} N.V.Krasnikov, Int.J.Mod.Phys. {\bf 22} 5117 (2007).
\bibitem{3} N.V.Krasnikov, Mod.Phys.Lett. {\bf A23} 3233 (2008).
\bibitem{4} A.A.Slavnov, Theor.Math.Phys.{\bf 148} 1159 (2006).
\bibitem{5} H.Nikolic, Mod.Phys.Lett. {\bf A23} 2645 (2008).
\bibitem{6} G.A.Kozlov, arXiv:0905.2272[hep-ph] (2009). 
\bibitem{7} H.Georgi, Phys.Rev.Lett. {\bf 98} 221601 (2007).
\bibitem{8} M.A.Stephanov, Phys.Rev. {\bf D76} 035008 (2007).

\end{thebibliography}
\end{document}